\documentclass{iau} 
\usepackage{graphicx}

\usepackage{xcolor}

\usepackage[normalem]{ulem}

\title[JD 11.~~Milky Way and Andromeda collision] 
{The collision between\\the Milky Way and Andromeda\\and the fate of their\\Supermassive Black Holes}

\author[R. Schiavi \& R. Capuzzo -Dolcetta \& M. Arca Sedda \& M. Spera]   
{Riccardo Schiavi$^1$, Roberto Capuzzo Dolcetta$^1$, Manuel Arca Sedda$^2$, Mario Spera$^3$ $^4$}

\affiliation{$^1$Department of Physics, Universit\`a La Sapienza,\\
Roma, 00185, Italy \\ email: {\tt riccardo.schiavi@uniroma1.it} \\[\affilskip]
$^2$ARI, University of Heidelberg, Heidelberg, 69117, Germany\\
$^3$Dep. of Physics and Astronomy, Universit\`a di Padova, Italy\\
$^4$Dep. of Physics and Astronomy, CIERA-Northwestern University\\}

\pubyear{2019}
\volume{351}  
\jname{Star Clusters: From the Milky Way to the Early Universe}
\editors{A. Bragaglia, M.B. Davies, A. Sills \& E. Vesperini, eds.}
\begin{document}

\maketitle

\begin{abstract}
Our Galaxy and the nearby Andromeda Galaxy (M31) form a bound system, even though the relative velocity vector of M31 is currently not well constrained. Their orbital motion is highly dependent on the initial conditions, but all the reliable scenarios imply a first close approach in the next 3$-$5 Gyrs. In our study, we simulate this interaction via direct $N$-body integration, using the HiGPUs code. Our aim is to investigate the dependence of the time of the merger on the physical and dynamical properties of the system. Finally, we study the dynamical evolution of the two Supermassive Black Holes placed in the two galactic centers, with the future aim to achieve a proper resolution to follow their motion until they form a tight binary system.
\end{abstract}

\firstsection 
\section{Introduction}

Although the physical and dynamical properties of the Milky Way-Andromeda system are rather uncertain, it is likely that the two galaxies will not escape the collision and the final merger. According to previous simulations in literature (\cite[Cox \& Loeb 2008]{cl2008}, \cite[Van der Marel 2019]{vdm2019}), 
the first close approach will likely occur in 5 Gyr from today and the final merger in about 10 Gyr; however the timing of this process is highly sensitive not only to the value of the tangential component of the M31's relative velocity, but also to the total mass of the binary and the density of the intergalactic medium (IGM). A clear understanding of how these properties affect the dynamics of the system will allow us to define a feasible scenario of the Local Group evolution. At the same time, during the interaction at large scale, we are interested in following the motion of the two SMBHs in the centers of both galaxies. After the merger of their host galaxies the two SMBHs are expected to form a close binary, which will gradually lose orbital energy during gravitational encounters with the field stars in a dense environment.
Our main purpose is to increase the resolution of our simulations to follow the dynamical evolution of the two SMBHs down to small spatial scales, so to deal with the `final parsec problem', when most of the orbital energy may be lost through gravitational waves (GW) emission.

\section{Initial conditions}

\begin{figure}
\begin{center}
\includegraphics[width=4.0in]{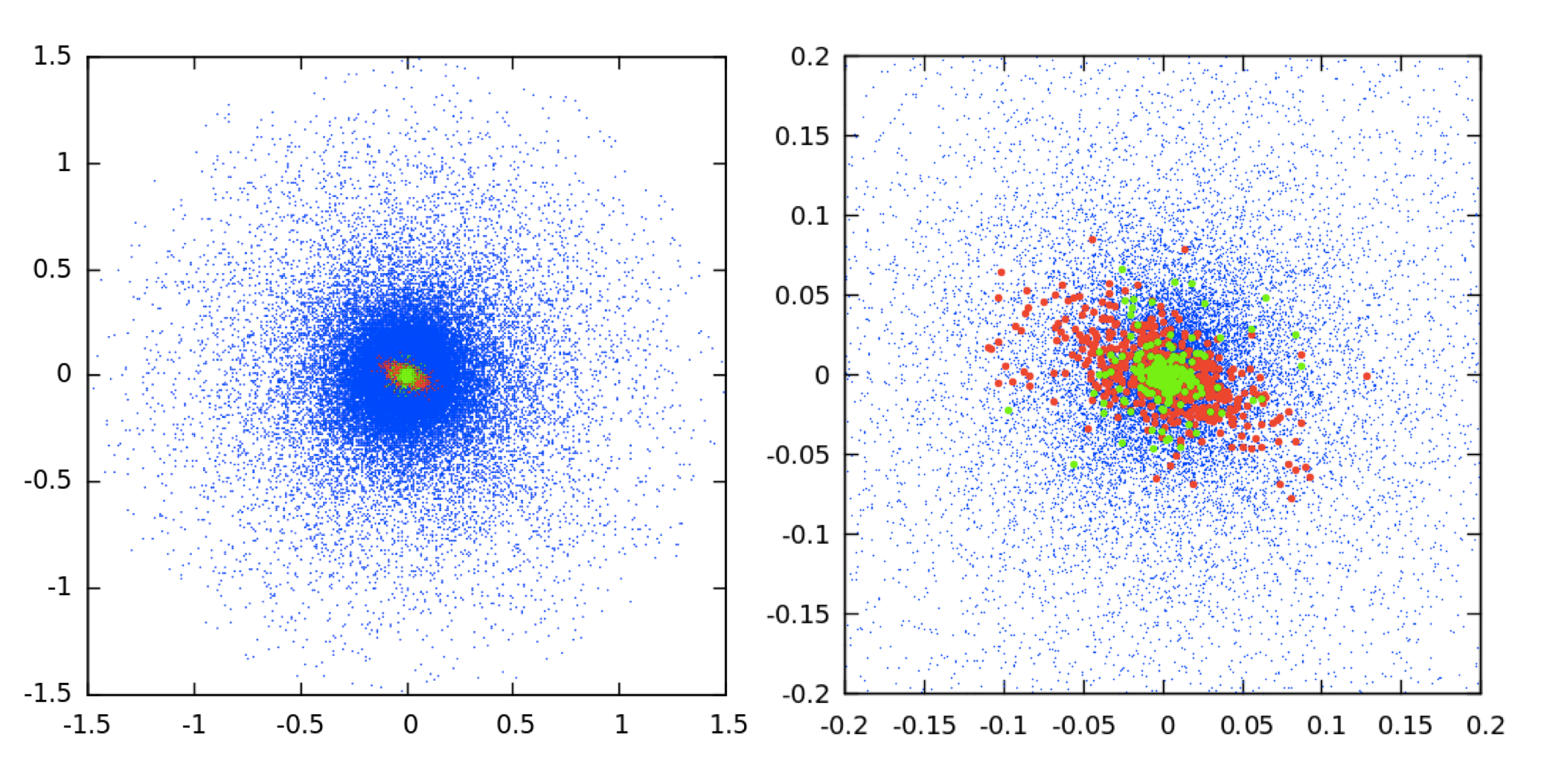} 
\caption{Our model of the Milky Way, with the three  components (disk, bulge, halo) shown in different colors. The right panel is a zoom of the innermost region. The axis unit is 100 kpc.}
\label{fig1}
\end{center}
\end{figure}

Initial conditions for our galaxies have been generated with the NEMO code (\cite[Teuben 1995]{Teuben1995}), combining three different components: a disk with an exponential density profile, a bulge and a halo, both with a Hernquist’s profile.
For both galaxies, the mass of the bulge is half the disk, and that of the halo about 42 times larger. In unit of disk's scale length, the core radius of the bulge and the halo are respectively 0.2 and 3 (\cite[Cox \& Loeb 2008]{cl2008}, \cite[Widrow \& Dubinski 2005]{Widrow-Dubinki}). Our simulations show a significant correlation between the cut-off radius of the two halos and the time of the interaction. For this reason we have chosen a large cut-off of 70 times the disk's scale length, which for the Milky Way corresponds to a radius of about 150 kpc, in line with the estimations of \cite[Shull (2014)]{Shull2014}. In Fig.\,\ref{fig1} we show a snapshot of our initial model for the Milky Way. \\
In the center of both galaxies we considered a black hole with a mass of one thousandth the total mass of the galaxy ($1.0\times10^{9} M_\odot$ for the Milky Way and $1.6\times10^{9}M_\odot$ for Andromeda). Even though these masses are different from the observed ones, until now this is the most reasonable choice, taking into account our current number resolution of $N=6.5\times10^4$ particles. \\
We have chosen 780 kpc as initial separation between the Milky Way and Andromeda centers of mass, with galaxy spin vectors respectively oriented at $(0^{\circ}; -90^{\circ})$ and $(241^{\circ}; -30^{\circ})$ in Galactic coordinates (\cite[Raychaudury  Lynden-Bell 1989]{R.L.B.}). In our simulations, we have adopted a reference frame where the x-y plane coincides with the plane of the motion.\\
The Andromeda's current velocity vector is one of the worst constrained property of the MW-M31 system. Using the redshift we can infer only the value of the radial component ($V_{r0}\approx120$ km/s), but measuring the transverse component is very difficult. Several estimations in literature give different values: from a minimum of $V_{t0}\approx17 km/s$ (\cite[Van der Marel 2012]{vdm2012}) to a maximum of $V_{t0}\approx164 km/s$ (\cite[Salomon 2016]{Sal2016}). Recently, \cite[Van der Marel et al. 2019]{VdM2019}, using the GAIA DR2 data, have obtained $V_{t0}\approx57$ km/s, but the level of uncertainty is still high. We took the latter estimation as reference for fixing the orientation of the M31's velocity vector, and we have performed three simulations varying transverse component speed ($30$ km/s, $50$ km/s, $70$ km/s), keeping constant the value of the radial part ($120$ km/s).

In Tab.\,\ref{tab1} we show all the physical properties of the two galaxies.

\begin{table}
 \begin{center}
 \caption{Values of characteristic parameters used in our simulations. Lengths are in kpc and masses in solar masses.}
  \label{tab1}
 {\scriptsize
  \begin{tabular}{|l|c|c|}\hline 
  & {\bf Milky Way} & {\bf Andromeda} \\ \hline
{\bf Scale radius of the disk ($R_d$)} & $2.2$ & $3.6$ \\ \hline
{\bf Core radius of the bulge(in unit of $R_d$)} & $0.2$ & $0.2$ \\ \hline
{\bf Core radius of the halo(in unit of $R_d$)} & $3.0$ & $3.0$ \\ \hline
{\bf Cut-off radius of the bulge(in unit of $R_d$)} & $5.0$ & $5.0$ \\ \hline
{\bf Cut-off radius of the halo(in unit of $R_d$)} & $70.0$ & $70.0$ \\ \hline
{\bf Mass of the disk ($M_d$)} & $2.27\times10^{10}$ & $3.64\times10^{10}$ \\ \hline
{\bf Mass of the bulge (in unit of $M_d$)} & $0.5$ & $0.5$ \\ \hline
{\bf Mass of the halo (in unit of $M_d$)} & $42.5$ & $42.5$ \\ \hline
{\bf Total Mass} & $1.0\times10^{12}$ & $1.6\times10^{12}$ \\ \hline
{\bf Mass of the central black hole} & $1.0\times10^{9}$ & $1.6\times10^{9}$ \\ \hline
  \end{tabular}
  }
 \end{center} 
 \end{table}

\section{Methods and Preliminary Results}
\begin{figure}
\begin{center}
\includegraphics[width=3.0in]{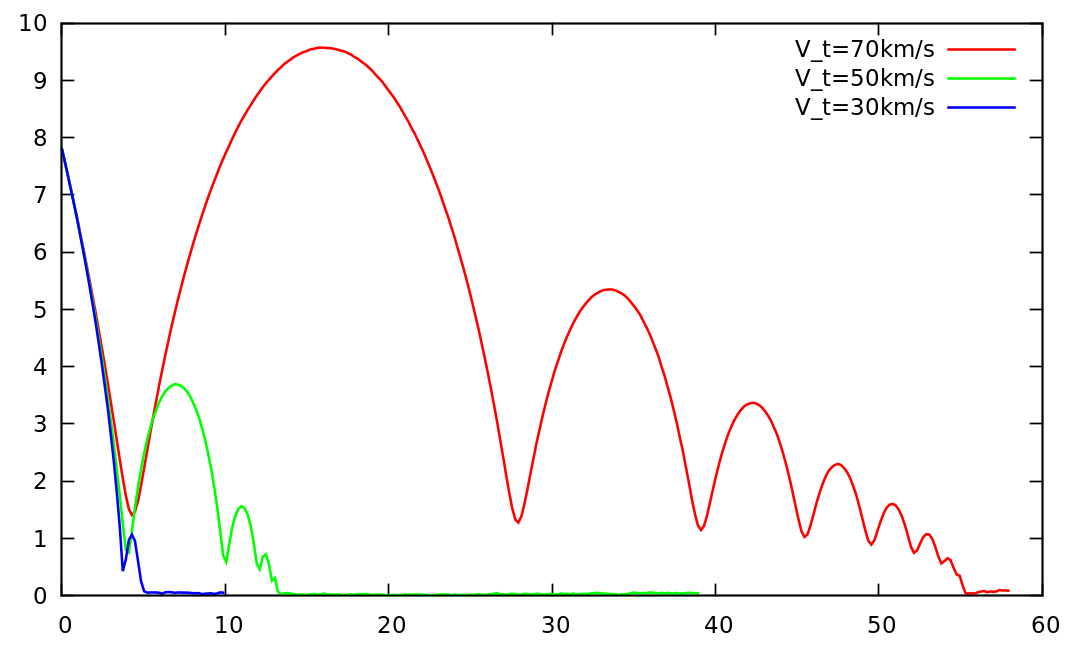} 
\caption{The separation (in unit of 100 kpc) between the two galaxies centers of mass as function of time (in Gyr), for three different tangential velocities}
\label{fig2}
\end{center}
\end{figure}

\begin{figure}
\begin{center}
\includegraphics[width=3.0in]{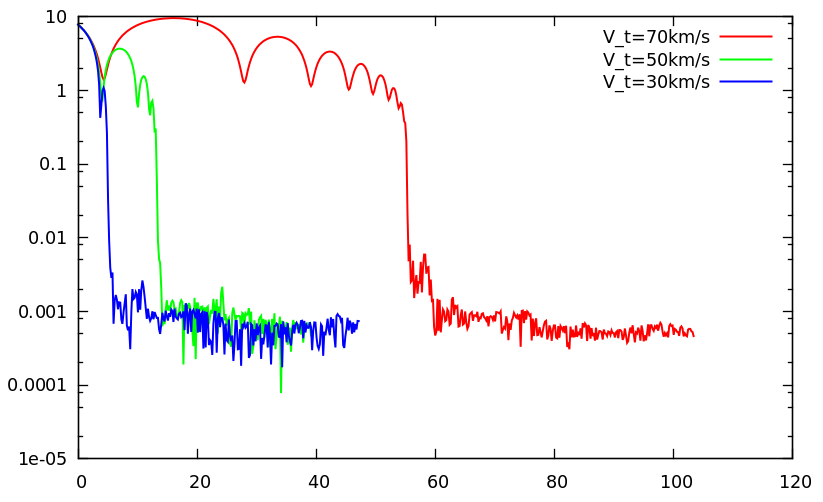} 
\caption{The separation (in unit of 100 kpc) between the central BHs as function of time (in Gyr), for three different tangential velocities. The plot shows that, regardless of the initial velocity, the two BHs stalls at the same  distance of about 50 pc.}
\label{fig3}
\end{center}
\end{figure}

The HiGPUs code (\cite[Capuzzo Dolcetta et al. 2013]{Capuzzo2013}), used in these simulations, numerically implements the integration of the classical $N$-body problem using CPUs coupled to GPUs, in parallel. This program is based on a 6th order Hermite’s integration scheme with block time steps, and directly evaluates the particle-particle forces. To reproduce the dissipative effect of the IGM, we have modified the original version of HiGPUs by including a dynamical friction term in the equations of motion, according to the Chandrasekhar's formula (\cite[Binney \& Tremaine 1987]{binney}). Anyway, our simulations indicates that the presence of a warm ($T\approx3\times10^5 K$) medium with a uniform density, $\rho_0$, which is roughly 10 times the critical density of the universe, produces just a secondary effect on the timescale of the interaction. The main factor driving the speed of the merger are the sizes of the galactic halos: the larger they are, the faster is the process.\\
Our first goal has been to investigate the dependence of the time of the merger upon the initial conditions, in particular upon the M31's initial transverse velocity. Figure\,\ref{fig2} shows the separation between the two galaxies as function of time, for the three chosen values of $V_{t}$, keeping the radial component fixed at $120$ km/s. The time of the merger, as expected, is highly sensitive to this parameter.\\
Looking at the relative motion of the two BHs, in Fig.\,\ref{fig3}, we notice that, independently of the magnitude and orientation of the initial relative velocity, they reach the same final separation of about $50$ pc, when they seem to stall on a nearly circular orbit. Anyway, our numerical resolution is too low to further investigate the BHs motion at small scales, because, clearly, the final value of their separation depends significantly on the number of particle in the simulation.

\section{Conclusions}
Owing to the uncertainty on the relative tangential velocity of Andromeda, the expected time of the MW-M31 merger spans over a wide range. Moreover, we have noticed that it is heavily dependent also on the radii of the galactic halos, and, secondarily, on the density and temperature of the IGM. We are currently performing other simulations to investigate further these dependencies, with the aim of a full description of the different scenarios of the galactic collision. 

The relative motion of the two central BHs shows a regular behaviour on large scales, but we need to increase the resolution of our simulations to extend the study also to the innermost region of the galaxy formed after the merger.

\end{document}